\documentclass[conference]{IEEEtran}
\IEEEoverridecommandlockouts
\usepackage{amsthm}
\usepackage{amsmath}
\usepackage{amssymb}
\usepackage{booktabs}
\usepackage{siunitx}
\usepackage{cite}
\usepackage{graphics, framed}
\usepackage{graphicx}
\usepackage{epsfig}
\usepackage{epstopdf}
\usepackage{subfigure}
\usepackage{url}
\usepackage{multimedia}
\usepackage{paralist}
\usepackage[printonlyused]{acronym}
\usepackage{units}
\usepackage{balance}
\usepackage{multirow}
\usepackage{color,soul}
\newcommand{\FGR}[1]{Fig.~\ref{#1}}

\newcommand{\SEC}[1]{Section~\ref{#1}}
\newcommand{\TAB}[1]{Table~\ref{#1}}

\renewcommand{\baselinestretch}{0.86}
\acrodef{5G}[5G]{5\textsuperscript{th}-Generation}
\acrodef{BW}[BW]{bandwidth}
\acrodef{BS}[BS]{base station}
\acrodef{CW}[CW]{continuous wave}
\acrodef{D2D}[D2D]{device-to-device}
\acrodef{dB}[dB]{decibel}
\acrodef{dBi}[dBi]{decibel isotropic}
\acrodef{dBm}[dBm]{decibel over a milliwatt}
\acrodef{GAL}[GAL]{graph attention layer}
\acrodef{GAT}[GAT]{graph attention network}
\acrodef{Gbps}[Gbps]{gigabit per second}
\acrodef{GHz}[GHz]{gigahertz}
\acrodef{THz}[THz]{Terahertz}
\acrodef{RIS}[RIS]{reconfigurable intelligent surface}
\acrodef{CN}[CN]{core network}
\acrodef{PSK}[PSK]{phase shift keying}
\acrodef{QAM}[QAM]{quadrature amplitude modulation}
\acrodef{AWGN}[AWGN]{additive white Gaussian noise}
\acrodef{SNR}[SNR]{signal-to-noise ratio}
\acrodef{AF}[AF]{amplitude-and-forward}
\acrodef{MIMO}[MIMO]{multiple-input multiple-output}
\acrodef{mMIMO}[mMIMO]{massive-multiple-input multiple-output}
\acrodef{SDN}[SDN]{Software-defined network}
\acrodef{SON}[SON]{self-organizing network}
\acrodef{hetnet}[HetNet]{heterogeneous network}
\acrodef{FSO}[FSO]{free-space optics}
\acrodef{UM-MIMO}[UM-MIMO]{ultra-massive-MIMO}
\acrodef{AP}[AP]{access point}
\acrodef{UE}[UE]{user equipment}
\acrodef{NFP}[NFP]{networked flying platform}
\acrodef{UAV}[UAV]{unmanned aerial vehicle}
\acrodef{HAPS}[HAPS]{high-altitude platform station}
\acrodef{LEO}[LEO]{low-earth orbit}
\acrodef{BAN}[BAN]{body area network}
\acrodef{WLAN}[WLAN]{wireless local area network}
\acrodef{QoS}[QoS]{quality of service}
\acrodef{TCS}[TCS]{thermal control system}
\acrodef{QCL}[QCL]{quantum cascade laser}
\acrodef{CMOS}[CMOS]{complementary metal-oxide semiconductor}
\acrodef{V-HetNet}[V-HetNet]{vertical heterogeneous network}
\acrodef{DL}[DL]{deep learning}
\acrodef{DRL}[DRL]{deep reinforcement learning}
\acrodef{FDTD}[FDTD]{Finite-difference time-domain}
\acrodef{FEM}[FEM]{finite element method}
\acrodef{MoM}[MoM]{method of moments}
\acrodef{VNA}[VNA]{vector network analyzer}
\acrodef{CS}[CS]{channel sounder}
\acrodef{CIR}[CIR]{channel impulse response}
\acrodef{CTF}[CTF]{channel transfer function}
\acrodef{PN}[PN]{pseudo-noise}
\acrodef{TOA}[TOA]{time of arrival}
\acrodef{GMM}[GMM]{Gaussian mixture model}
\acrodef{OOK}[OOK]{on-off keying}
\acrodef{MLE}[MLE]{maximum likelihood estimation}
\acrodef{LOS}[LOS]{line-of-sight}
\acrodef{NLOS}[NLOS]{non-line-of-sight}
\acrodef{SG}[SG]{signal generator}
\acrodef{SA}[SA]{spectrum analyzer}
\acrodef{FDSOI}[FDSOI]{fully depleted silicon on insulator}
\acrodef{OpEx}[OpEx]{operational expenditures}
\acrodef{TDD}[TDD]{time division duplex}
\acrodef{CSI}[CSI]{channel state information}
\acrodef{MAC}[MAC]{medium access control}
\acrodef{GEO}[GEO]{geostationary orbit}
\acrodef{SWaP}[SWaP]{size, weight, and power}
\acrodef{NMSE}[NMSE]{normalized mean square error}
\acrodef{MSE}[MSE]{mean square error}
\acrodef{LS}[LS]{least square}
\ifCLASSINFOpdf
\else
\fi

\begin{document}

\title{Channel Estimation for \textcolor{black}{Full-Duplex} RIS-assisted HAPS Backhauling with Graph Attention Networks}
\IEEEoverridecommandlockouts 

\author{\IEEEauthorblockN{K{\"{u}}r{\c{s}}at~Tekb{\i}y{\i}k\IEEEauthorrefmark{1}, G{\"{u}}ne{\c{s}}~Karabulut~Kurt\IEEEauthorrefmark{1}, Chongwen~Huang\IEEEauthorrefmark{2},\\ Ali~R{\i}za~Ekti\IEEEauthorrefmark{3}, Halim~Yanikomeroglu\IEEEauthorrefmark{4}}

\IEEEauthorblockA{\IEEEauthorrefmark{1}Department of Electronics and Communication Engineering, {\.{I}}stanbul Technical University, {\.{I}}stanbul, Turkey}

\IEEEauthorblockA{\IEEEauthorrefmark{2}College of Information Science and Electronic Engineering, Zhejiang University, Hangzhou, China}

\IEEEauthorblockA{\IEEEauthorrefmark{3}Department of Electrical--Electronics Engineering, Bal{{\i}}kesir University, Bal{{\i}}kesir, Turkey}

\IEEEauthorblockA{\IEEEauthorrefmark{4}Department of Systems and Computer Engineering, Carleton University, Ottawa, ON, Canada\\ Emails: \texttt{\{tekbiyik, gkurt\}@itu.edu.tr,} \texttt{chongwenhuang@zju.edu.cn,}\\ \texttt{arekti@balikesir.edu.tr,} \texttt{halim@sce.carleton.ca}}}

\maketitle

\begin{abstract}
In this paper, \ac{GAT} is firstly utilized for the channel estimation. \textcolor{black}{In accordance with the 6G expectations, we consider a \ac{HAPS} mounted} reconfigurable intelligent surface-assisted two-way communications \textcolor{black}{and obtain} a low overhead and a high \acl{NMSE} performance. The performance of the proposed method is investigated on the two-way backhauling link over the RIS-integrated \ac{HAPS}. The simulation results denote that the \ac{GAT} estimator overperforms the \acl{LS} in full-duplex channel estimation. Contrary to the previously introduced methods, \ac{GAT} at one of the nodes can separately estimate the cascaded channel coefficients. Thus, there is no need to use \acl{TDD} mode during pilot signaling in full-duplex communication. Moreover, it is shown that the \ac{GAT} estimator is robust to hardware imperfections and changes in small scale fading characteristics even if the training data do not include all these variations. 
\end{abstract}
\IEEEpeerreviewmaketitle
\acresetall

\section{Introduction}

As the state-of-the-art reflective surface, \acp{RIS} pave the way for a low-cost promising technology improving the spectral efficiency and energy efficiency in wireless networks~\cite{basar2019wireless, huang2019reconfigurable}. Since \acp{RIS} are able to manipulate the amplitude and/or phase of the impinging signal, they can substantially nullify the randomness of the propagation medium. However, to do these, a \ac{RIS}-assisted wireless communication system strictly requires high-quality \ac{CSI}. Moreover, channel estimation is getting much more complex for \ac{RIS}-assisted two-way communications, which has been proposed in~\cite{atapattu2020reconfigurable, zhang2020sum, shen2020beamforming} recently. Also, since the coefficients of channels (i.e., \ac{CN} to \ac{RIS} and \ac{RIS} to \ac{BS}) must be acquired for each channel induced by meta-atoms, the increasing number of elements significantly raises the overhead of the communication and leads to decrease in the efficiency. This study focuses on developing a channel estimation method for \ac{RIS}-assisted full-duplex communications without the need to activate \ac{TDD} mode during channel estimation as well as providing lightweight overhead.

\textcolor{black}{In this study, we will focus on a 6G compliant scenario, specifically look into \ac{V-HetNet} architecture. The \ac{V-HetNet}  is an emerging network topology including geostationary and low-earth orbit satellites, and \ac{HAPS} systems along with terrestrial communication links to serve a large number of small cells with the goals of ubiquitous connectivity and user-centric communication~\cite{tekbiyik2020holistic}. The \ac{HAPS} systems,  network nodes operating at an altitude of about $20$ km in the stratosphere, are the key enablers of the \acp{V-HetNet}. Due to the properties of the stratosphere, a \ac{HAPS} can remain in an almost stationary position and can provide ubiquitous connectivity~\cite{HighAltionline}. One of the main uses of \ac{HAPS} systems is to serve backhauling due to the high cost of fiber optic infrastructure~\cite{alzenad2018fso, kurt_vision_2020}. Many reflective surfaces can be deployed on a \ac{HAPS} thanks to their large surface area in order to enable  high capacity communication~\cite{alfattani_link_2020}.}

\textcolor{black}{Considering the variations in the channel characteristics observed by the receiving nodes in \ac{HAPS} communication due to scatterers, and clouds, the cascaded channel estimation in full-duplex \ac{RIS}-assisted \ac{HAPS} communication is challenging and requires the novel approaches which are robust to changes in the channel characteristics as well as hardware impairments. Below, we provide a solution to this problem by proposing the \ac{GAT} channel estimator.}

\subsection{Related Works}

A few works (e.g., \cite{atapattu2020reconfigurable, zhang2020sum, shen2020beamforming}) investigate \ac{RIS}-assisted two-way communications. In \cite{atapattu2020reconfigurable}, the phase shifts are optimized to maximize the minimum \ac{SNR}. The phase shifts and source precoders are jointly optimized in order to maximize the system sum rate in full-duplex \ac{MIMO} communications~\cite{zhang2020sum, shen2020beamforming}. They reveal the upper bound performances of the proposed systems because they assume that channels are perfectly predicted, which is not the case in practice. One should note that channels need to be estimated precisely in order to perform self-interference cancellation in full-duplex communication. The majority of channel estimation methods are not suitable for \ac{RIS}-assisted full-duplex communications owing to their high computational cost and/or expense of more overhead. 

The prominent channel estimation methods proposed for unidirectional \ac{RIS}-assisted communications are mentioned below, and the main drawbacks that are possibly experienced during channel estimation even in \ac{TDD} mode are discussed. Some of the existing methods assume that only a single meta-atom is active at a given time period~\cite{mishra2019channel, elbir2020deep}. This method is called on-off state control, which is time-consuming for a massive number of meta-atoms and allows utilizing the small portion of the elements due to a few active elements at a time even though utilizes \ac{DL}~\cite{elbir2020deep}. Furthermore, this method hampers the synchronization of meta-atoms since the channel coefficients probably change when the last one is estimated. Therefore, the synchronization is spoiled by this switching delay. At this point, an extra algorithm or method is required to recover the synchronization error between meta-atoms. Obviously, this new recovery algorithm would increase the computational complexity and time delay. As \acp{RIS} are composed of massive passive scattering elements, they are not able to estimate the channel coefficients on their own. Some methods, such as given by~\cite{taha2019enabling}, propose channel estimation using \acp{RIS} with active elements at the cost of destroying this attractive feature of \acp{RIS}. In this approach, \acp{RIS} are used with active sensors, which are equipped with baseband signal processing units for channel estimation. On the other hand, the channel estimation methods proposed in~\cite{wei2020channel} with massive element \acp{RIS} do not seem feasible since their computational complexity is proportional to the cube of the number of \ac{RIS} elements (i.e., $\mathcal{O}(N^3)$). Frankly, these methods are not capable to acquire both channel states in \acp{RIS}-assisted full-duplex communications. To the best knowledge of the authors, there is not any channel estimation method directly focusing on full-duplex communications, yet. Furthermore, unlike existing deep learning methods, we incorporate the \acp{GAT} to minimize the computational complexity~\cite{velickovic_graph_2018} and increase the learning rate by handling unseen nodes within the proposed \ac{RIS}-assisted full-duplex communication system.

Even though fiber optic communications are conventionally utilized in backhaul connectivity, building fiber optic infrastructure for small cells is mostly costly solution~\cite{dahrouj2015cost}. Recently, \ac{HAPS} systems have been proposed to deal with the cost of backhaul connectivities~\cite{kurt_vision_2020}. Therefore, we consider a two-way \ac{RIS}-assisted \ac{HAPS} backhauling by considering the prominent features of \ac{HAPS} backhauling~\cite{karapantazis2005broadband} and \ac{RIS}-enhanced two-way communications~\cite{atapattu2020reconfigurable} for an example application of the proposed \ac{GAT} channel estimator. 

\subsection{Contributions}

The main contributions of this study are two-fold and can be summarized as follows:
\begin{itemize}
    \item To the best of the authors' knowledge, this study firstly investigates a channel estimation method performing at only one of the nodes which transmit their data over full-duplex wireless communication links in order to acquire the coefficients for both channels as illustrated in \FGR{fig:system_model}. In virtue of the proposed channel estimation method, there is no need to switch half-duplex instead of full-duplex when estimating channels. Another crucial point is that this method does not need an on-off state control, which is a time-consuming approach in the channel estimation method proposed in the previous studies owing to the capability of acquiring all channel coefficients regarding the elements of \ac{RIS}.
    
    \item This study considers \ac{GAT} in channel estimation (even in wireless communications) for the first time. Owing to the attention mechanism in \ac{GAT}, the proposed system is robust against changes in channel parameters even if the network is trained under different and better channel conditions.
\end{itemize}
Note that the proposed method can be easily revised to be used for multi-user and/or \ac{MIMO} \ac{RIS}-assisted wireless communication scenarios. For instance, it is thought that it may be sufficient to expand the label vector to include channels related to each antenna for channel estimation in \ac{RIS}-assisted \ac{MIMO}. It should be highlighted that the method can be also utilized in half-duplex systems by limiting the label vector with only single channel coefficients. \textcolor{black}{Moreover, it should be emphasized that the proposed channel estimation method can be also applied to the channel estimation problems other than \ac{RIS}-assisted \ac{HAPS} communications.}

\subsection{Outline}
The remainder of this work first addresses the mathematical background and channel estimation for \ac{RIS}-assisted full-duplex communication in~\SEC{sec:system_model}. \SEC{sec:dl_model} provides the basic mathematical background for \acp{GAT}. Next, in~\SEC{sec:methodology}, the \ac{GAT} channel estimator  and the channel estimation methodology are introduced. \SEC{sec:results} presents the numerical and simulation results with discussions regarding the proposed \ac{GAT} estimator and \ac{LS}. Finally, \SEC{sec:conclusion} concludes this study.


\section{\ac{RIS}-assisted Full-Duplex Communications}\label{sec:system_model}

\begin{figure}[!t]
    \centering
    \includegraphics[width=\columnwidth]{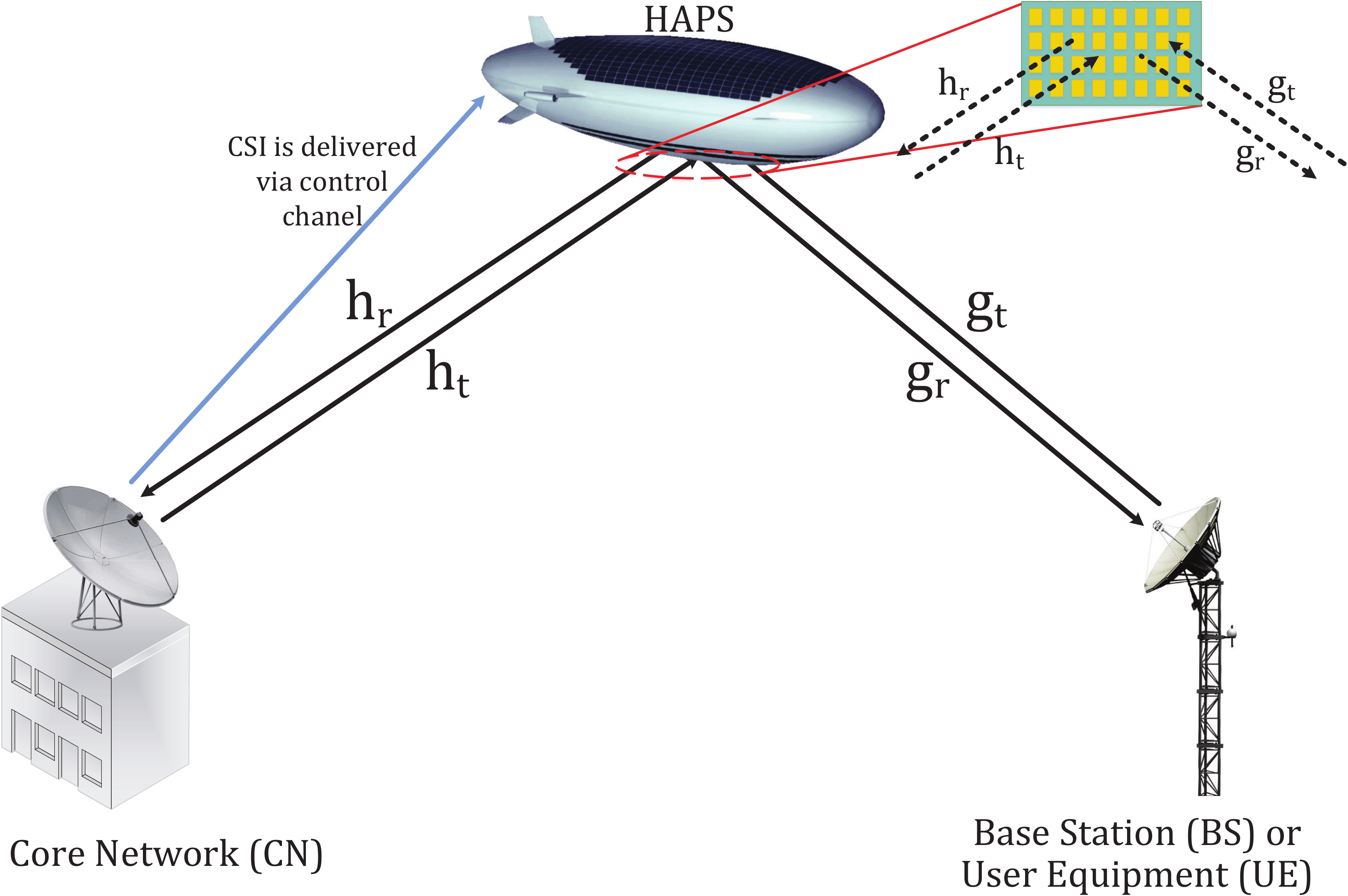}
    \caption{Full-duplex backhaul communication from \ac{BS} to \ac{CN} supported by \ac{RIS}-assisted \ac{HAPS}.}
    \label{fig:system_model}
\end{figure}

A considered two-way backhaul network including two end nodes, namely \ac{CN} and a \ac{BS}\footnote{This node can also be considered as a user equipment.}, supported by a \ac{RIS}-assisted \ac{HAPS} is denoted in \FGR{fig:system_model}. Each end node transmits their information at the same time via the \ac{RIS} which has $N$ passive elements. Due to long range between nodes and possible obstacles, it can be assumed that the direct link between nodes is negligibly weak. Therefore, the direct link is ignored in this study. Furthermore, the channels \ac{CN}-to-\ac{RIS} and \ac{RIS}-to-\ac{BS} can be assumed reciprocal if the nodes transmit within a coherence interval and the antennas on the nodes are placed closely. Under the reciprocity assumption, it can be said that $\mathbf{h_t} = \mathbf{h_r} = \mathbf{h}$ and $\mathbf{g_t} = \mathbf{g_r} = \mathbf{g}$, where $\mathbf{h}$ and $\mathbf{g}$ are channel coefficient vectors such that $\mathbf{h} = \left[h_1, h_2, \dots, h_N\right]$ $\mathbf{g} = \left[g_1, g_2, \dots, g_N\right]$. Each fading coefficient for the wireless channel between \ac{CN} and the $n$-th element of \ac{RIS} is denoted $h_n = \alpha_n \mathrm{e}^{-j \varphi_n}$. Similarly, the channel coefficient for the second channel is $g_n = \beta_n \mathrm{e}^{-j \psi_n}$. For both ground-to-\ac{HAPS} and \ac{HAPS}-to-ground channels, the magnitudes of channel coefficients can be assumed to follow the Rician distribution~\cite{michailidis2010three}. As each node transmits concurrently, the received signal at the node \ac{CN} can be given as
\begin{align}
y_{1}(t)=& \sqrt{P_{2}}\left(\sum_{n=1}^{N} g_{n} \kappa\mathrm{e}^{j(\theta_{n}-\psi_{n}-\varphi_{n})} h_{n}\right) s_{2}(t) + e_{1}(t) \nonumber \\ 
&+\sqrt{P_{1}}\left(\sum_{n=1}^{L} h_{n} \kappa\mathrm{e}^{j(\theta_{n}-2\varphi_{n})} h_{n}\right) s_{1}(t) + w_{1}(t),
\label{eq:rec_signal}
\end{align}
where the first term is desired signal; $\kappa$ and $\theta_{n}$ denote amplitude gain and the adjustable phase shift at $n$-th element of \ac{RIS}, respectively. $\sqrt{P_1}$ and $\sqrt{P_2}$ stand for the received powers of the \ac{CN} and \ac{BS}, respectively. The nodes \ac{CN} and \ac{BS} simultaneously transmit symbols $s_1(t)$ and $s_2(t)$, which are the pilot symbols in this study. $e_{1}(t)$ and $w_{1}(t)$ are the residual loop interference and \ac{AWGN} at \ac{CN}; can be assumed to be distributed with $\mathcal{C} \mathcal{N}\left(0, \sigma_{e_{1}}^{2}\right)$~\cite{rodriguez2014performance} and $\mathcal{C} \mathcal{N}\left(0, \sigma_{w_{1}}^{2}\right)$, respectively. For the simplicity, the phase shifts created by the \ac{RIS} can be denoted with a diagonal matrix, $\mathbf{\Theta} \in \mathbb{C}^{N\times N}$, as
\begin{align}
    \mathbf{\Theta} = \mathrm{diag} \left\{\kappa\mathrm{e}^{j\theta_1}, \dots, \kappa\mathrm{e}^{j\theta_N} \right\}.
\end{align}
Then, (\ref{eq:rec_signal}) can be rewritten as
\begin{align}
y_{1}(t)=\sqrt{P_{2}} \mathbf{h}^{\mathrm{T}} \mathbf{\Theta g} s_{2}(t)+\sqrt{P_{1}} \mathbf{h}^{\mathrm{T}} \boldsymbol{\Theta} \mathbf{h} s_{1}(t)+e_{1}(t)+w_{1}(t),
\end{align}
where $\mathbf{h}^{\mathrm{T}} \boldsymbol{\Theta} \mathbf{h} s_{1}(t)$ is the self-interference. Similar to $y_1(t)$, the received signal at \ac{BS} is given as
\begin{align}
y_{2}(t)=\sqrt{P_{1}} \mathbf{g}^{\mathrm{T}} \mathbf{\Theta h} s_{1}(t)+\sqrt{P_{2}} \mathbf{g}^{\mathrm{T}} \boldsymbol{\Theta} \mathbf{g} s_{2}(t)+e_{2}(t)+w_{2}(t).
\end{align}

To express the received signal in terms of channel estimation terminology better, the received pilot symbols at \ac{CN} for the transmitted pilot symbols, $\mathbf{s}_{1} \in \mathbb{C}^{1\times M}$, can be restated as follows
\begin{align}
    \mathbf{y_{1}} = \sqrt{P_{2}} \mathbf{g \Theta h} \mathbf{s_{2}} + \sqrt{P_{1}} \mathbf{h}^{\mathrm{T}} \boldsymbol{\Theta} \mathbf{h} \mathbf{s_{1}} + \mathbf{e_{1}} + \mathbf{w_{1}},
\end{align}
where, $\mathbf{y_{1}},\, \mathbf{e_{1}},\, \mathbf{w_{1}} \in \mathbb{C}^{1\times M}$ and $\mathbf{h^{\mathrm{T}}}, \mathbf{g^{\mathrm{T}}} \in \mathbb{C}^{1\times N}$. Regarding the received pilot symbols, the instantaneous \ac{SNR} related to the estimation of $\mathbf{h}$ at \ac{CN} can be given as \begin{align}
    \mathbf{\gamma_{1,h}} = \frac{|\sqrt{P_{2}} \mathbf{g \Theta h} \mathbf{s_{2}}|^{\circ2}}{|\mathbf{e_{1}}|^{\circ2} + |\mathbf{w_{1}}|^{\circ2}},
\end{align}
where $(\cdot)^{\circ2}$ denotes the Hadamard square. The \ac{LS} estimation for $\mathbf{h}$ is defined as 
\begin{align}
    \hat{\mathbf{h}} = \left(\mathbf{S}^{\mathrm{T}}\mathbf{S}\right)^{-1}\mathbf{S}^{\mathrm{T}}\mathbf{y_1}^{\mathrm{T}},
\end{align}
where
\begin{align}
    \mathbf{S} = \begin{bmatrix}
s_1(1) & s_1(1) & \dots & s_1(1)\\
\vdots & \ddots & & \vdots\\
s_1(M) & s_1(M) & \dots & s_1(M)\\
\end{bmatrix}.
\end{align}
\ac{NMSE} is given as 
\begin{align}
\mathrm{NMSE} = \frac{\|\mathbf{h}-\hat{\mathbf{h}}\|_{2}^{2}}{\|\mathbf{h}\|_{2}^{2}},
\end{align}
where $\|\cdot\|_{2}$ represents the Euclidean norm.


\section{The Graph Attention Networks}\label{sec:dl_model}

While conventional \ac{DL} methods such as convolutional neural networks are successful on data that exhibit a grid-like structure, they cannot show high performance on data in the irregular domain. Graph neural networks can produce state-of-the-art solutions for problems involving data that do not have a grid-like structure. In addition, the attention mechanism provides a very suitable structure for inductive learning, so that the trained network can be generalized over unobserved graphs. Therefore, \acp{GAT} can be considered as an emerging healer in channel estimation where the observed data frequently changes because of the random nature of the propagation medium. The mathematical background of \acp{GAT} is detailed below.

A set of $P$ nodes which is an input to the \ac{GAL} is defined as $\mathbf{\vartheta}=\left\{\vec{\vartheta}_{1}, \vec{\vartheta}_{2}, \ldots, \vec{\vartheta}_{P}\right\}, \, \vec{\vartheta}_{i} \in \mathbb{R}^{F}$, where $F$ denotes the number of features for each node. \ac{GAL} outputs a set of node features as $\mathbf{\vartheta}^{\prime}=\left\{\vec{\vartheta}_{1}^{\prime}, \vec{\vartheta}_{2}^{\prime}, \ldots, \vec{\vartheta}_{P}^{\prime}\right\}, \, \vec{\vartheta}_{i}^{\prime} \in \mathbb{R}^{F^{\prime}}$, where $F^{\prime}$ denotes the number of features for each output nodes and might have different cardinality than the former. A shared linear transformation parameterized by the weight matrix, $\mathbf{W} \in \mathbb{R}^{F\times F^{\prime}}$, is applied to each node to transform input properties to higher-level properties. Thereafter, the self-attention on the nodes is investigated by an attention mechanism $a: \mathbb{R}^{F^{\prime}} \times \mathbb{R}^{F^{\prime}} \rightarrow \mathbb{R}$, which computes the attention coefficients as
\begin{align}
c_{i j}=a\left(\mathbf{W} \vec{\vartheta}_{i}, \mathbf{W} \vec{\vartheta}_{j}\right),
\end{align}
where $c_{i j}$ is found for only the $j$-th node which has neighborhood of $i$-th node in the graph. The attention coefficients show the importance of features of $j$-th node on the $i$-th node. To make the coefficients comparable in a different neighborhood, they are then normalized using the softmax function as follows~\cite{bahdanau_neural_2016}
\begin{align}
\alpha_{i j}=\operatorname{softmax}_{j}\left(c_{i j}\right)=\frac{\exp \left(c_{i j}\right)}{\sum_{k \in \mathcal{N}_{i}} \exp \left(c_{i k}\right)},
\end{align}
where $\mathcal{N}_{i}$ denotes the neighborhood of the $i$-th node. The normalized coefficients, $\alpha_{i j}$, are computed by the attention mechanism $a$ as~\cite{velickovic_graph_2018}
\begin{equation}
\alpha_{i j}=\frac{\exp \left(\operatorname{ReLU}\left(\mathbf{a}^{\top}\left[(\mathbf{X} \mathbf{W})_{i} \|(\mathbf{X} \mathbf{W})_{j}\right]\right)\right)}{\sum_{k \in \mathcal{N}(i)} \exp \left(\operatorname{ReLU}\left(\mathbf{a}^{\top}\left[(\mathbf{X} \mathbf{W})_{i} \|(\mathbf{X} \mathbf{W})_{k}\right]\right)\right)},
\end{equation}
where $\mathbf{a} \in \mathbb{R}^{2 F^{\prime}}$ and $\mathbf{X} \in \mathbb{R}^{P \times F}$ denote attention kernel and node features, respectively. Finally, the convolution over the graph network is performed as 
\begin{equation}
\mathbf{Z}=\alpha \mathbf{X} \mathbf{W}+\mathbf{b},
\end{equation}
where $\mathbf{b}$ is the trainable bias vector. This layer accepts inputs which are the node attributes matrix $\mathbf{X} \in \mathbb{R}^{P \times F}$, binary adjacency matrix $\mathbf{A} \in\{0,1\}^{P \times P}$, and edge attributes matrix $\mathbf{E} \in \mathbb{R}^{P \times P \times S}$.

As stated in~\cite{lee_self-attention_2019}, a pooling layer is required to generalize graph convolution networks. Furthermore, to reduce the number of representations, a pooling layer is employed. Thus, the pooling layer enables the network to avoid overfitting. Since this study utilizes global attention pooling, we do not refer to other graph pooling layers here. The global attention pooling layer computes
\begin{align}
\mathbf{X}^{\prime}=\sum_{i=1}^{P}\left(\sigma\left(\mathbf{X} \mathbf{W}_{1}+\mathbf{b}_{1}\right) \odot\left(\mathbf{X} \mathbf{W}_{2}+\mathbf{b}_{2}\right)\right)_{i},
\end{align}
where $\sigma$ and $\odot$ are the sigmoid function and the broadcasted elementwise product, respectively.

\section{Full-Duplex Channel Estimation}\label{sec:methodology}

\begin{figure}[!t]
    \centering
    \includegraphics[width=\columnwidth]{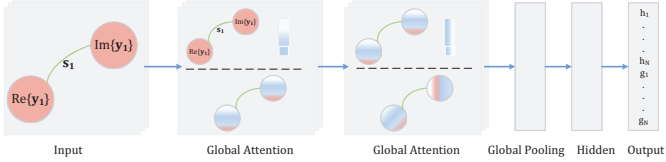}
    \caption{The illustration for the proposed \acl{GAT} including two consecutive \acl{GAT}s and global attention pooling. The real and imaginary parts of received signal, $\mathbf{y_1}$, are assigned to attributes of two nodes. The edge attributes are set as the pilot symbols, $\mathbf{s_1}$.}
    \label{fig:graph}
\end{figure}

\begin{table}[]
\centering
\caption{Summary of the dataset parameters employed during training and test.}
\begin{tabular}{lcc}
\toprule \toprule
\multicolumn{1}{l}{\textbf{Parameters}} & \multicolumn{1}{c}{\textbf{Training}} & \multicolumn{1}{c}{\textbf{Test}} \\  \midrule
\multicolumn{1}{l}{PN Polynomial}       & $x^4+x^2+1$                           & $x^4+x^2+1$                       \\
\multicolumn{1}{l}{Modulation}       & BPSK                           & BPSK                       \\
\multicolumn{1}{l}{\# of Samples per SNR}       & 1000                                  & 500                               \\
\multicolumn{1}{l}{SNR (dB)}            & -30:2:0                               & -30:2:10                          \\
\multicolumn{1}{l}{K}                   & 10                                    & 0, 4, 8, 10, 12                   \\
\multicolumn{1}{l}{$M$}                   & 16, 32, 64, 128                                    & 16, 32, 64, 128                   \\
\multicolumn{1}{l}{$N$}                   & 128, 256, 512, 1024                                    & 128, 256, 512, 1024                   \\
\multicolumn{1}{l}{$\epsilon$}          & 0                                     & 0, 1e-1, 1e-2, 1e-3               \\ \bottomrule \bottomrule
\end{tabular}
\label{tab:parameters}
\end{table}

In this section, we introduce a channel estimation procedure using \ac{GAT} for two-way backhaul over \ac{RIS}-assisted \ac{HAPS} as illustrated in \FGR{fig:system_model}. As aforementioned, it is required that channel estimation should be carried out without switching \ac{RIS} elements on-off and be robust against hardware imperfections and serious fluctuations in channel characteristics. \acp{GAT} can be generalized to completely unobserved graphs during the training~\cite{velickovic_graph_2018}; therefore, it is an appropriate solution to the channel estimation problem. Considering the random nature of the wireless propagation medium and that \ac{RIS} cannot manipulate random behavior during the channel estimation, a generalizable model or procedure is required for accurate channel estimation in order to avoid performance degradation in estimation when channel characteristics significantly change. In consequence, \acp{GAT} are utilized in this study to obtain all channel parameters regarding \ac{RIS} elements in a one-shot that is to say without consecutive on-off switch for each element on \ac{RIS}.

\subsection{Training Dataset Generation}\label{sec:dataset_generation}

Instead of using \ac{TDD} mode during pilot signaling, we consider remaining two-way communications when channel estimation is performed. Hence, we can define the problem as estimation of $\mathbf{h}$ and $\mathbf{g}$ from $\mathbf{y_1}$ at node \ac{CN}. Similary, the same problem can be defined for the node \ac{BS}; however, we consider only $\mathbf{y_1}$ for the estimation of $\mathbf{h}$ and $\mathbf{g}$. 

Firstly, the $M$-length pilot symbols are created as \ac{PN} sequence utilizing the polynomial given by $x^4+x^2+1$. Utilizing \ac{PN} sequence also enables the nodes and \ac{HAPS} to make synchronous, which is demanded by a proper wireless communications. In this study, $\mathbf{s_1}$ and $\mathbf{s_2}$ are selected same. Then, this pilot symbols are received at \ac{CN} after reflected from \ac{RIS} with $N$ meta-atoms. $P_1$ and $P_2$ are chosen as unit power. During pilot signaling, all elements of \ac{RIS} can switch on without any phase shift, scilicet unitary phase shift matrix, $\boldsymbol{\Theta}$. In other word, $\kappa = 1$. As the scatterers which are close to ground stations are required to be considered~\cite{dovis2002small}, we assume that both channels depicted in \FGR{fig:system_model} follow Rician fading with $K = 10$. By using these parameters, the training dataset is generated. The input regarding received signal, $\mathbf{X}$, is created as follows:
\begin{align}
    \mathbf{X} &= \left[\mathrm{Re}\{\mathbf{y_1}\}; \, \mathrm{Im}\{\mathbf{y_1}\}\right].
\end{align}
The adjacency matrix for the graph network denoting the real and imaginary parts of channel coefficient is given as
\begin{align}
\mathbf{A} = \begin{bmatrix}
0 & 1 \\
1 & 0
\end{bmatrix},
\end{align}
which implies that two nodes are connected with a single edge as illustrated in \FGR{fig:graph}. The known pilot symbols are assigned to weight of the edge for the $k$-th nonzero element of adjacency matrix as
\begin{align}
\mathbf{E_k} = \mathbf{s_{1}}, \, \mathbf{E} \in \mathbb{C}^{2\times2\times M}.
\end{align}
Also, the label vector related to these inputs is defined as
\begin{align}
\mathbf{y} = \left[h_1, h_2, \cdots, h_N, g_1, g_2, \cdots, g_N\right]^{\mathrm{T}}.
\end{align}
The training dataset includes $1000$ input samples for each \ac{SNR} value in between -$30$ dB and $0$ dB with $2$ dB step. The chosen \ac{SNR} interval allows proper \ac{RIS}-assisted communications in terms of bit error rate as seen in~\cite{basar2019wireless}. The training dataset totally consists of $16000$ input samples for each $M$, $N$, and \ac{SNR} values. The dataset is divided into two parts, which are used for training and validation with the rate of $4:1$. The parameters which are used during the training dataset generation are summarized in \TAB{tab:parameters}.

\begin{table}[!t]
\centering
\caption{The proposed \ac{GAT} layout and parameters.}
\begin{tabular}{ccc}
\toprule \toprule
\multicolumn{2}{c}{\textbf{Layers}}   & \textbf{Dimensions}  \\ \midrule
\multirow{3}{*}{Inputs}      & $\mathbf{X}$      & $2\times M$          \\ \cmidrule{2-3}
                             & $\mathbf{A}$      & $2\times 2$          \\ \cmidrule{2-3}
                             & $\mathbf{E}$      & $2\times 2 \times M$ \\ \midrule
\multirow{1}{*}{Labels}      & $\mathbf{y}$      & $2N \times 1$   \\ \midrule             
\multicolumn{2}{c}{Graph Attention 1} & $2\times 128$        \\
\multicolumn{2}{c}{Graph Attention 2} & $2\times 32$         \\
\multicolumn{2}{c}{Global Attention Pool}   & 128                  \\
\multicolumn{2}{c}{Dense}             & $2N$    \\ \toprule

\multicolumn{2}{c}{\textbf{Parameters}}   & \textbf{Values}  \\ \midrule
\multicolumn{2}{c}{Activation} & ReLU  \\
\multicolumn{2}{c}{Optimizer} & Adam  \\
\multicolumn{2}{c}{Loss} & MSE  \\
\multicolumn{2}{c}{Learning Rate} & 1e-3  \\
\multicolumn{2}{c}{L$_2$ Regularization} & 5e-4  \\
   
\bottomrule   \bottomrule 
\end{tabular}
\label{tab:gat_layout}
\end{table}

\subsection{\ac{GAT} Model and Training}
In \SEC{sec:dl_model}, the background for \acp{GAT} is visited. In this section, we introduce the parameters and details related to the proposed \ac{GAT} model. Two consecutive \acp{GAL} are used to learn the channel parameters over the graph structure detailed before. The first one has $128$ output channels while the latter has $32$. In two layers, ReLU activation function is employed. Considering the dataset regarding channel estimation problem, the dimensions of inputs now become as $P = 2$, $F = M$, and $S = M$. Besides \acp{GAL}, a global attention pooling layer is employed in order to reduce the number of representations and thus avoid the network overfitting. To keep away the network from overfitting problem, the dropout in both \acp{GAL} with the rate of $0.5$ and $\mathrm{L}_2$ regularization are utilized. The network is terminated by a hidden layer with $2N$ neurons. The optimizer and loss function are chosen as ADAM with the learning rate of $10^{-3}$ and \ac{MSE}. We utilize Spektral~\cite{grattarola2020graph} in the implementation of graph neural network, namely \ac{GAT}. The number of epochs is set to $20$, but early stopping is activated if there is no improvement in the loss for $5$ epochs. The parameters for the \ac{GAT} and inputs are summarized in \TAB{tab:gat_layout}.

\begin{figure}[!t]
    \centering
    \includegraphics[width=\columnwidth]{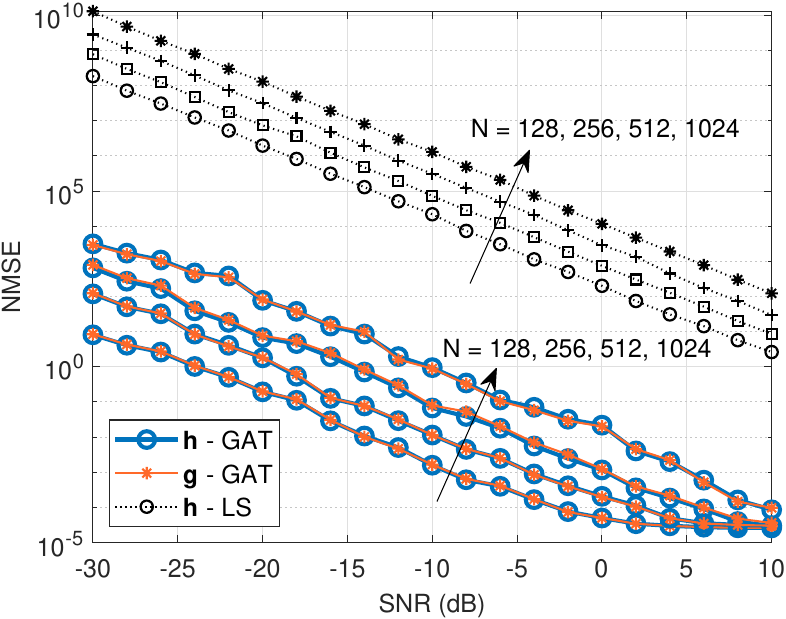}
    \caption{\ac{NMSE} performance of the proposed \ac{GAT}-aided full-duplex channel estimation versus the \acp{SNR} and the number of \ac{RIS} elements, $N$, for $M = 128$, $K = 10$ and $\epsilon = 0$. The proposed method can estimate the channel coefficients, $\mathbf{h}$ and $\mathbf{g}$, with almost the same performance at \ac{CN}.}
    \label{fig:h_g_m_128}
\end{figure}

\begin{figure}[!t]
    \centering
    \includegraphics[width=\columnwidth]{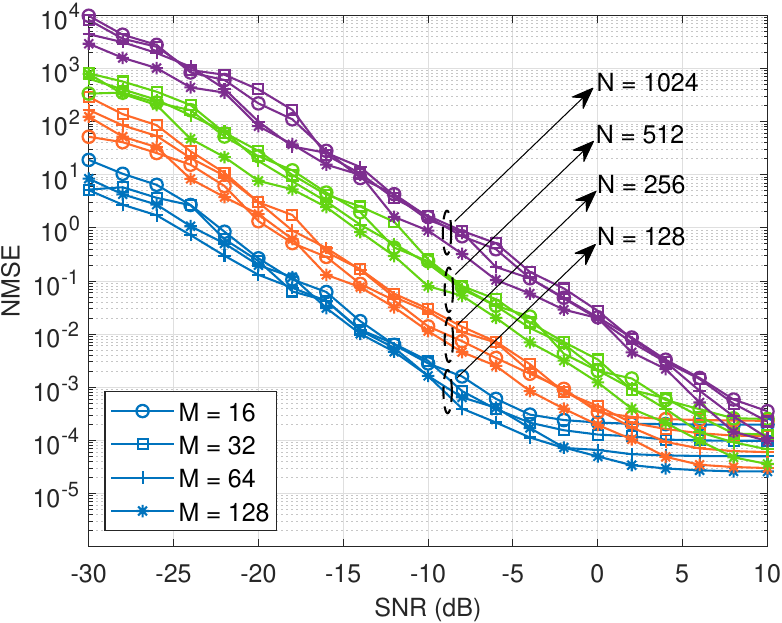}
    \caption{\ac{NMSE} performance of the proposed \ac{GAT}-aided full-duplex channel estimation versus the \acp{SNR}, the number of \ac{RIS} elements, $N$, and the number of pilot symbols, $M$, for $K = 10$ and $\epsilon = 0$.}
    \label{fig:g_all}
\end{figure}

\begin{figure}[!t]
    \centering
    \includegraphics[width=\columnwidth]{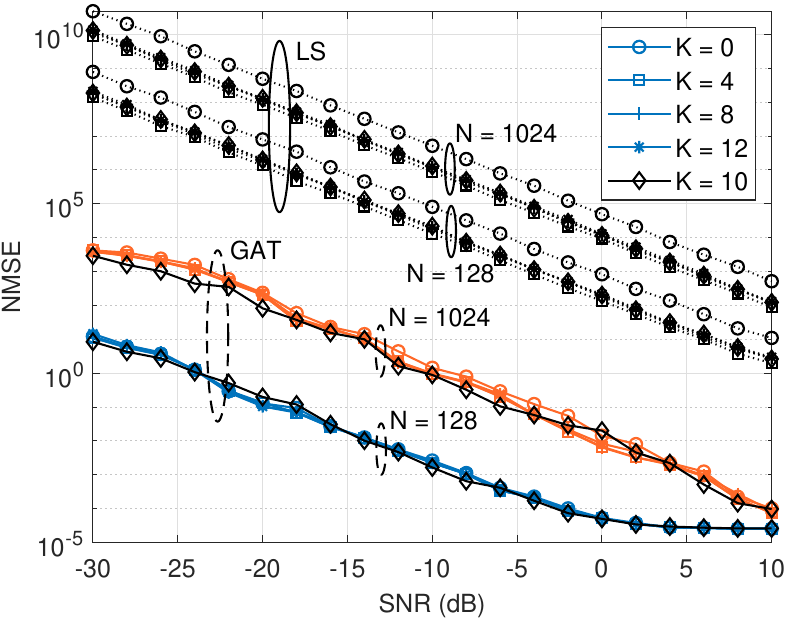}
    \caption{\ac{NMSE} performance of the proposed \ac{GAT}-aided full-duplex channel estimation versus the \acp{SNR}, the number of \ac{RIS} elements, $N$, and the Rician $K$-factor for $M = 128$ and $\epsilon = 0$. The proposed method is able to remain almost the same performance under the effect of changing fading.}
    \label{fig:g_all_K}
\end{figure}

\begin{figure}[!t]
    \centering
    \includegraphics[width=\columnwidth]{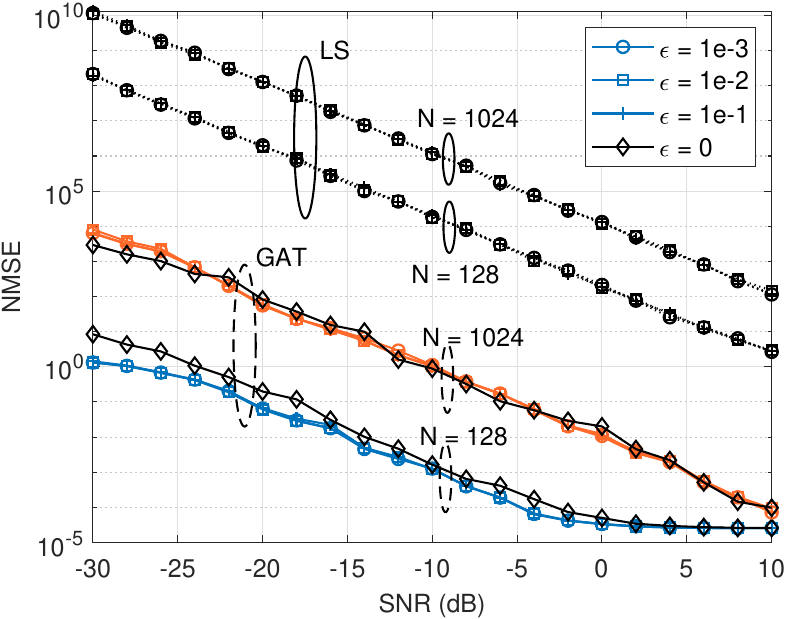}
    \caption{\ac{NMSE} performance of the proposed \ac{GAT}-aided full-duplex channel estimation versus the \acp{SNR}, the number of \ac{RIS} elements, $N$, and switching error, $\epsilon$, for $M = 128$ and $K = 10$. The proposed method is able to remain almost the same performance under the effect of changing hardware imperfection.}
    \label{fig:g_all_EPS}
\end{figure}



\section{Numerical Results and Discussions}\label{sec:results}

In this section, the channel estimation performance of the \ac{GAT} model, the details of which are given in the previous section, is investigated under different scenarios. Firstly, we consider the full-duplex channel estimation performances for both $\mathbf{h}$ and $\mathbf{g}$ in \FGR{fig:h_g_m_128}. The proposed \ac{GAT} estimator outperforms \ac{LS} estimation where \ac{TDD} is not activated. Both methods show the same \ac{NMSE} performance with about $3$ dB of \ac{SNR} loss when the number of \ac{RIS} elements is doubled and the number of pilot symbols is kept constant (i.e. $M = 128$). Although the estimation procedure is carried out at only \ac{CN}\footnote{The same \ac{GAT} can be used at \ac{BS} without any change in the parameters and architecture.}, \FGR{fig:h_g_m_128} denotes almost the same performance for both channels, which shows that the main purpose of this study is successfully achieved. Since \ac{NMSE} performance for two channel estimations are almost identical, only \ac{NMSE} results for $\mathbf{h}$ are given hereafter to increase the readability of the figures. It should be noted that although \ac{SNR} values between $0$ and $10$ dB are not included in the training set, the proposed method can make channel estimation with high performance at these \ac{SNR} values.

Also, we investigate the channel estimation performance jointly for the number of pilot symbols and the number of \ac{RIS} elements as seen in \FGR{fig:g_all}. It is observed that the increase in the number of pilot symbols does not make a significant difference in \ac{NMSE} performance except for the maximum achievable performance limit at the high \ac{SNR} region. The main reason behind these results is that the attention mechanism used in this \ac{DL} network can focus on the most relevant part of inputs when making a decision, as stated in~\cite{velickovic_graph_2018}. In the light of the results, it can be said that the \ac{GAT} estimator can reduce the overhead for channel estimation. On the other hand, increasing the number of channel coefficients required to be estimated slightly deteriorates the \ac{NMSE} performance. As said before, when the number of meta-atoms is doubled, the system needs $3$ dB more \ac{SNR} to remain the same \ac{NMSE} performance.

As the value of $K$ is dependent on the propagation environment, for instance, it can be observed that $K$ is lower in urbanized regions where many scatterers are usually found (vice versa in rural areas). Furthermore, it should be noted that the \ac{HAPS} movement or displacement can give rise to change in the platform elevation angle; thereby, the value of $K$ changes with the variation in the elevation angle~\cite{michailidis2010three}. In consequence, it is important to test the performance of the channel estimation method according to the small scale fading factor. In~\FGR{fig:g_all_K}, the trained \ac{GAT} model is tested under varying $K$ values even though the network is trained with only $K = 10$. While the \ac{LS} estimator's performance degrades in the case that there is no line-of-sight link, \ac{GAT} estimator is able to show satisfactory \ac{NMSE} performance regardless of $K$ values. This is another finding that \acp{GAT} can be successful in different inputs.

 Although the reflection coefficients of \ac{RIS} elements are assumed as $1$, which means perfect reflector, it should be noted that they cannot completely reflect the power of an incident wave. Thus, we consider the imperfection in the amplitude gain of \acp{RIS} by introducing an error, $\varepsilon$, such that $\kappa = 1- \epsilon$ as in~\cite{elbir2020deep}. The network which is trained under the ideal amplitude condition is tested for $\epsilon = 10^{-3}, \, 10^{-2}, \, 10^{-1}$. As expected, \FGR{fig:g_all_EPS} presents that the \ac{LS} estimator is not affected from the error in amplitude gain. The \ac{GAT} provides much better channel estimation performance than \ac{LS}. Moreover, its performance is not degraded when changing the amplitude gain of \ac{RIS}. To speak generally, \FGR{fig:g_all_EPS} denotes the robustness of the proposed channel estimation method against the hardware imperfections.

The attention mechanism in graph convolutional networks, pooling, and regularization used in the proposed \ac{GAT} channel estimator help ensure robustness and high performance as shown in the results above. Noted that the proposed system can be utilized for the channel estimation in half-duplex communications as well. For example, $\mathbf{g}$ and $\mathbf{h}$ are separately estimated at \ac{CN} and \ac{BS}, respectively when \ac{TDD} mode is activated.


\section{Concluding Remarks}\label{sec:conclusion}
 In this study, we propose a channel estimation method utilizing \ac{GAT} which can be generalized over unobserved graph structures in the virtue of the attention mechanism. The proposed method does not require on-off state control and can estimate the coefficients of two main channel blocks of the RIS-assisted communication system at the same time with a high performance. \ac{GAT} estimator's \ac{NMSE} performance is studied in the scenario full-duplex backhauling over \ac{RIS}-integrated \ac{HAPS}. 
 
 The simulation results reveal that the proposed method has remarkably high performance. Furthermore, thanks to the attention mechanism and graph structure, the estimator is able to maintain its performance under different channel conditions and hardware impairments which are not seen by the network during the training. Besides full-duplex channel estimation, the \ac{GAT} estimator can also be employed in half-duplex communications and multi-user systems as well as \ac{MIMO} systems by modifying the label vector in the training phase.

\balance
\bibliographystyle{IEEEtran}
\bibliography{icc_haps_ris}
\end{document}